\documentclass[graybox]{svmult}
\usepackage{mathptmx}       
\usepackage{helvet}         
\usepackage{courier}        
\usepackage{type1cm}        
\usepackage{makeidx}         
\usepackage{graphicx}        
\usepackage{multicol}        
\usepackage[bottom]{footmisc}

\usepackage{amsmath}

\newcommand{\dna}{{\mathord{\downarrow}}}
\newcommand{\upa}{{\mathord{\uparrow}}}

\usepackage{hyperref}
\usepackage{url}

\newcommand{\expval}[1]{\left< #1 \right>}
\newcommand{\ket}[1]{\left|#1\right.\rangle}
\newcommand{\bra}[1]{\langle\left.#1\right|}

\newcommand{\nn}{\nonumber\\}
\newcommand{\f}[1]{\mbox{\boldmath$#1$}}

\newcommand{\ord}{{\cal O}}

\newcommand{\traceB}[1]{{\rm Tr_B}\left\{ #1 \right\}}

\newcommand{\abs}[1]{{\left| #1 \right|}}

\newcommand{\ii}{\mathrm{i}}  

\begin{document}
  
\title{How small can Maxwell's demon be?}
\subtitle{Lessons from autonomous electronic feedback models}

\author{Gernot Schaller}

\institute{Gernot Schaller \at Helmholtz-Zentrum Dresden-Rossendorf, Bautzner Landstra{\ss}e 400, 01328 Dresden, Germany,\\
\email{g.schaller@hzdr.de}}
%
%

\maketitle

\abstract{External piecewise-constant feedback control can modify energetic and entropic balances, allowing in extreme scenarios 
for Maxwell demon operational modes.
Without specifying the actual implementation of external feedback loops, one can only partially quantify the additional contributions to entropy production.
This is different in autonomously operating systems with internal feedback.
Traditional (bipartite) autonomous systems can be divided into controller and a controlled subsystem, but also non-bipartite systems can accomplish the same task.
We consider examples of autonomous three-terminal models that transfer heat mainly from a cold to a hot reservoir by 
dumping a small fraction of it to an ultra-cold (demon) reservoir, such that their coarse-grained dynamics resembles an external feedback loop.  
We find that the minimal three-level implementation is most efficient in utilizing heat dissipation to change the entropy balance of the effective controlled system.
}


Maxwell's demon is a hypothetical being that in the original gedanken experiment~\cite{maxwell1871} monitors the speed and direction of thermal particles and uses that information to spatially separate the particles into faster and slower fractions by opening and closing a shutter (which ideally requires no energy expenditure).
This paradox has fascinated generations of researchers: Entropy is (locally) reduced and the resulting gradient could even be used to extract useful work, such that ideally the demon uses only information to lead to this apparent violation of the second law of thermodynamics.
Formally speaking, it implements an external feedback control loop: A measurement reveals information about the particle (is it e.g. fast or slow, left or right-moving), the signal is then processed and the corresponding control operation is performed (shall the demon open or close the shutter).

Nowadays, Maxwell's paradox is resolved by including the entropic balance of the demon in the discussion~\cite{leff1990,leff2002,maruyama2009a}.
This already enters in effective descriptions of feedback control loops~\cite{esposito2012a,bergli2013a} where the information gained from measurements can be experimentally quantified~\cite{vidrighin2016a,cottet2017a}.
Alternatively, one can represent the demon by a separate physical system~\cite{mandal2013a,strasberg2013a} that is in contact with the controlled one and study the energy and information flows within the bipartite joint system~\cite{horowitz2014a,ehrich2023a}.
It has been argued that the processing of information in this traditional sense is not a necessary ingredient~\cite{sanchez2019a,ciliberto2020a}, which has fostered attempts to classify Maxwell demons more rigorously~\cite{freitas2021a}.

In any case, the actual nature of the feedback loop implementation becomes relevant, and in this book chapter contribution, we will compare how external feedback loops affect the energetic and entropic balance of a discrete (quantum) system with the situation in small autonomous systems that accomplish the same task.

\section{Background: Energy and Particle-resolving rate equations}

The dynamics of many open systems with discrete quantum states can be described by rate equations -- first order differential equations for probabilities $P_a$ to occupy state $a$ with energy $E_a$ and particle number $N_a$ -- of the form
\begin{align}\label{EQ:rate}
\dot P_a = \sum_\nu \sum_b R_{ab}^\nu P_b - \sum_\nu \sum_b R_{ba}^\nu P_a\,,
\end{align}
where $R_{ab}^\nu\ge 0$ is the transition rate from state $b$ to a different state $a$ due to reservoir $\nu$ (for stationary transport scenarios we require multiple reservoirs).
Naturally, given a proper initial state, \eqref{EQ:rate} obeys $P_a \in [0,1]$ and $\sum_a P_a = 1$ at all times~\footnote{ 
This is fulfilled when $R_{ab}^\nu \ge 0$ for $a \neq b$ and $R_{aa}^\nu=0$ (these diagonal values would cancel out anyways).
If we rewrite~\eqref{EQ:rate} in matrix form $\f{P} = {\cal R} \f{P}$, the off-diagonal elements of the rate matrix will be given by the rates ${\cal R}_{a \neq b} = \sum_\nu R_{ab}^\nu \ge 0$, whereas the diagonal entries are such that the column sum of the rate matrix vanishes ${\cal R}_{aa} = -\sum_\nu \sum_{b\neq a} R_{ba}^\nu$, sufficient for probability conservation.}.

Above, we have assumed that the transition rates are additively composed from different reservoirs.
Such rate equations may arise as a special case of Lindblad master equations~\cite{breuer2002} and have their (limited) regime of validity, typically based on
weak-coupling, Markovian and sometimes secular approximations.
One should therefore keep in mind that with this formalism, results for the ultra-short regime, the strong-coupling regime, and the regime of near degenerate energies should be treated with caution, as the exact dynamics in these regimes may be much more complicated (e.g. non-Markovian or having non-additive rates).
Furthermore, in the natural setting where one obtains these rate equations, one also assumes the reservoirs to be maintained at a thermal state, described by inverse temperature $\beta_\nu = (k_{\rm B} T_\nu)^{-1}$ and possibly a chemical potential $\mu_\nu$.
This eventually reflects in the fact that the transition rates for the same reservoir obey local detailed balance relations
\begin{align}\label{EQ:local_detailed_balance}
\frac{R_{ab}^\nu}{R_{ba}^\nu} = e^{-\beta_\nu[E_a-E_b-\mu_\nu(N_a-N_b)]}\,,
\end{align}
such that for $\mu_\nu=0$ the energy-increasing transitions are suppressed compared to energy-decreasing transitions, and raising the chemical potentials tends to increase the system particle number.
If the system is only coupled to a single reservoir or -- equivalently -- if all reservoirs are at the same inverse temperature $\beta$ and chemical potential $\mu$ (such that we can drop the bath index $\nu$), 
this can be used to show that the thermal state $\bar P_a = e^{-\beta (E_a - \mu N_a)}/Z$ is a stationary one~\footnote{
Consider $\sum_b R_{ab} \bar P_b - \sum_b R_{ba} \bar P_a = \sum_b R_{ab}/Z \left[e^{-\beta (E_b-\mu N_b)} - e^{+\beta[E_a-E_b-\mu(N_a-N_b)]} e^{-\beta(E_a-\mu N_a)}\right] = 0$.
For reservoirs at different equilibrium states, the stationary state is a non-equilibrium one.}.

\subsection{System energy and particle balance}

The time derivative of the system energy $\expval{E} = \sum_a E_a P_a$ can in absence of direct driving of the system be written as
\begin{align}
  \frac{d}{dt} \expval{E} = \sum_a E_a \dot P_a = \sum_\nu \sum_{ab} \left[E_a R_{ab}^\nu P_b - E_a R_{ba}^\nu P_a\right] 
  = \sum_\nu \sum_{ab} (E_a-E_b) R_{ab}^\nu P_b\,,
\end{align}
where (by the additivity of the transition rates) the time-dependent energy current entering the system from reservoir $\nu$ becomes
\begin{align}\label{EQ:cur_energy}
I_E^{\nu} \equiv \sum_{ab} (E_a-E_b) R_{ab}^\nu P_b\,.
\end{align}
Neglecting the energy content in the system-reservoir interaction (weak-coupling limit), we can identify the reservoir changes with the negative energy current entering the system $dE_\nu/dt \approx - I_E^\nu$, and then the above just implies the first law of thermodynamics
\begin{align}\label{EQ:first_law}
\frac{d}{dt} \expval{E} + \sum_\nu \frac{dE_\nu}{dt} =0\,,
\end{align}
indicating that the total energy of system and reservoirs together remains constant~\footnote{
Adding and subtracting the matter currents~\eqref{EQ:cur_matter} we can also write the first law~\eqref{EQ:first_law} as 
$\frac{d}{dt} \expval{E} = \sum_\nu \mu_\nu I_M^\nu + \sum_\nu (I_E^\nu - \mu_\nu I_M^\nu)$ in terms of a chemical work rate and heat currents, respectively.
}.
An analogous thought allows to construct the particle (matter) currents from the system particle number $\expval{N} = \sum_a N_a P_a$ via 
$\frac{d}{dt} \expval{N} \equiv \sum_\nu I_M^\nu$ with the time-dependent matter current entering the system from reservoir $\nu$ given by
\begin{align}\label{EQ:cur_matter}
I_M^{\nu}\equiv \sum_\nu \sum_{ab} (N_a-N_b) R_{ab}^\nu P_b\,.
\end{align}

To obtain the stationary currents $\bar I_{E/M}^\nu$ (we denote stationary values by an overbar throughout), one simply has to insert the stationary occupation probabilities for the system $P_b \to \bar P_b$.
For a single thermal reservoir obeying~\eqref{EQ:local_detailed_balance}, it follows that all currents at steady state must vanish, but this is no longer true for different reservoirs held at different temperatures and/or chemical potentials.

\subsection{System entropy balance}

The time derivative of the system's Shannon entropy  $\expval{S} = -\sum_a P_a \ln P_a$ becomes~\footnote{In the topmost line of~\eqref{EQ:entropy_balance} we used $\sum_a \dot P_a = 0$, inserted the rate equation~\eqref{EQ:rate} and expanded the fraction in the first term of the r.h.s.
In the second line we used properties of the logarithm and exchanged $a\leftrightarrow b$ in the last term.
In the third line, we have combined the logarithms to insert the local detailed balance property~\eqref{EQ:local_detailed_balance} and identified the currents~\eqref{EQ:cur_energy} and~\eqref{EQ:cur_matter} in the last line.}
\begin{align}\label{EQ:entropy_balance}
\frac{d}{dt} \expval{S} &= -\sum_a \dot P_a \ln P_a = -\sum_\nu \sum_{ab} R_{ab}^\nu P_b \ln P_a \frac{R_{ba}^\nu}{P_b R_{ab}^\nu} \frac{P_b R_{ab}^\nu}{R_{ba}^\nu}
+ \sum_\nu \sum_{ab} R_{ba}^\nu P_a \ln P_a\nn
&= +\sum_\nu \sum_{ab} R_{ab}^\nu P_b \ln \frac{R_{ab}^\nu P_b}{R_{ba}^\nu P_a} +\sum_\nu \sum_{ab} R_{ab}^\nu P_b \ln \frac{R_{ba}^\nu}{R_{ab}^\nu} \frac{1}{P_b} + \sum_\nu \sum_{ab} R_{ab}^\nu P_b \ln P_b\nn
&= +\sum_\nu \sum_{ab} R_{ab}^\nu P_b \ln \frac{R_{ab}^\nu P_b}{R_{ba}^\nu P_a} +\sum_\nu \sum_{ab} R_{ab}^\nu P_b \ln \frac{R_{ba}^\nu}{R_{ab}^\nu}\nn
&= +\sum_\nu \sum_{ab} R_{ab}^\nu P_b \ln \frac{R_{ab}^\nu P_b}{R_{ba}^\nu P_a} + \sum_\nu \beta_\nu (I_E^\nu -  \mu_\nu I_M^\nu)\,.
\end{align}
The first term on the r.h.s. of the last line is positive (which e.g. follows from the logarithmic sum inequality) and is known as (irreversible) entropy production rate~\cite{esposito2007a,strasberg2022}
\begin{align}
\dot\sigma_\ii \equiv +\sum_\nu \sum_{ab} R_{ab}^\nu P_b \ln \frac{R_{ab}^\nu P_b}{R_{ba}^\nu P_a} \ge 0\,.
\end{align}
The positivity of the entropy production rate quantifies the second law of thermodynamics 
\begin{align}\label{EQ:second_law}
\dot\sigma_\ii = \frac{d}{dt} \expval{S} - \sum_\nu \beta_\nu (I_E^\nu -  \mu_\nu I_M^\nu) = \frac{d}{dt} \expval{S} + \sum_\nu \frac{dS_\nu}{dt} \ge 0\,,
\end{align}
where we have identified the second term on the r.h.s. with the entropy production rate of a reservoir without volume change maintained in thermal equilibrium~\footnote{To see that, solve $dE_\nu = T_\nu dS_\nu + \mu_\nu dN_\nu$ for the reservoir entropy change $d S_\nu/dt$  and use that for weak couplings we have $dE_\nu/dt \approx -I_E^\nu$ and $dN_\nu/dt=-I_M^\nu$.}.

\subsection{Stochastic propagation}\label{SEC:stochastic_propagation}

To solve~\eqref{EQ:rate}, one may exponentiate the rate matrix ${\cal R}$, which is feasible when the dimension of ${\cal R}$ is small.
Such a solution would only provide average quantities, whereas a system subject to continuous measurements of its energy and/or particle number will fluctuate, such that the rate equation solution $P_i(t)$ has to be reconstructed from many experimental trajectories. 
Such time-resolved measurements can be more directly compared to stochastic trajectory solutions.

These are generated as follows: 
Supposing that at time $t$, the system is in state $a$, one determines the random waiting time $\tau_a$ until the next quantum jump via
\begin{align}\label{EQ:waiting_time}
\tau_a = \frac{-\ln(1-r)}{\sum_\nu \sum_{b\neq a} R_{ba}^\nu}\,,
\end{align}
where $r$ is a uniformly distributed random number with $r\in[0,1]$.
The waiting times determined this way are Poissonian-distributed according to the waiting-time distribution 
$\omega_a(\tau) = \sum_\nu \sum_{b\neq a} R_{ba}^\nu e^{-\sum_\nu \sum_{b\neq a} R_{ba}^\nu\tau}$.
The system remains in the state $a$ during $[t,t+\tau_a]$, and afterwards a quantum jump to a different state $b\neq a$ triggered by reservoir $\nu$ is performed with probability
\begin{align}\label{EQ:prob_jump}
P_{a\to b}^\nu = \frac{R_{ba}^\nu}{\sum_\mu \sum_{c\neq a} R_{ca}^\mu}\qquad:\qquad
\sum_{b\neq a} \sum_\nu P_{a\to b}^\nu=1\,,
\end{align}
such that after the waiting time, a jump is performed with certainty.
The energetic and particle changes are attributed to reservoir $\nu$, which allows to track the exchanged heat with the methods of stochastic thermodynamics~\cite{seifert2012a,strasberg2022}.
Then, one repeats the process with $b$ being the new initial state~\footnote{
Numerically, the algorithm has the advantage that no rate matrix needs to be stored, no timestep width needs to be adjusted, just random numbers are required.
To see that upon averaging trajectories one reproduces~\eqref{EQ:rate}, consider that the waiting time distribution according to~\eqref{EQ:waiting_time} just reproduces 
the probability to remain in state $a$ via
$P_{a\to a}(\tau) = 1 - \int_0^\tau \omega_a(t') dt' = e^{-\sum_\nu \sum_{b\neq a} R_{ba}^\nu\tau}$,  
and that the jump probability~\eqref{EQ:prob_jump} is just the probability for jump $a\to b$ compliant with~\eqref{EQ:rate}, conditioned on a jump actually happening.}.

Additional continuous measurements of the system energy and particle number will have no additional effect on the dynamics~\footnote{ 
This is only true if the measurement basis coincides with the pointer basis of the rate equation, e.g., if we measure the total system energy $E$ and total system particle number $N$.
If we measure e.g. local particle numbers, the competition of different bases may induce additional quantum features like the quantum Zeno effect~\cite{schaller2018a}.
}.

\subsection{Feedback control}\label{SEC:feedback_control}

The trajectory formulation can be directly generalized to immediate and piecewise-constant external interventions conditioned on a perfect measurement of the systems state or its transitions:
We assume that the control interventions only change the transition rates (e.g. via changing the system energies or the coupling strength to the reservoirs $\nu$), but not the eigenstates of the quantum system themselves (which would require a treatment beyond rate equations~\cite{schaller2018a}).
The quality of this assumption depends of course on the applied experimental feedback protocol.
Furthermore, we assume that the feedback operations are performed with negligible delay after the measurement (see e.g. Ref.~\cite{emary2013a} for delay effects) and that the control operations themselves leave the system particle number constant. 
Whereas ideal Maxwell-demon control leaves the system energy constant, more realistic control operations may change it~\cite{ehrich2023b}, and this feedback work has to be taken into account~\footnote{
As the feedback actions are conditioned on measurement results, the work rate for open-loop drivings $\dot W_{\rm drv} = \sum_a \dot E_a P_a$ does not apply here.}.

By convention, we label the energies that one would observe at a fixed moment in time (i.e., between the jumps) by $E_a$.
These may differ from the energies in absence of feedback, as e.g. upon detection of the final state $a$, the control operation has established energy $E_a$. 
Accordingly, the energy difference $E_a-E_b$ for a system's jump between two states $a$ and $b$ may now be decomposed into heat, stochastically exchanged with the reservoirs, and feedback work, which is tied to the stochastic heat contributions, and this decomposition can be done in different ways.

\begin{figure}
\includegraphics[width=0.8\textwidth,clip=true]{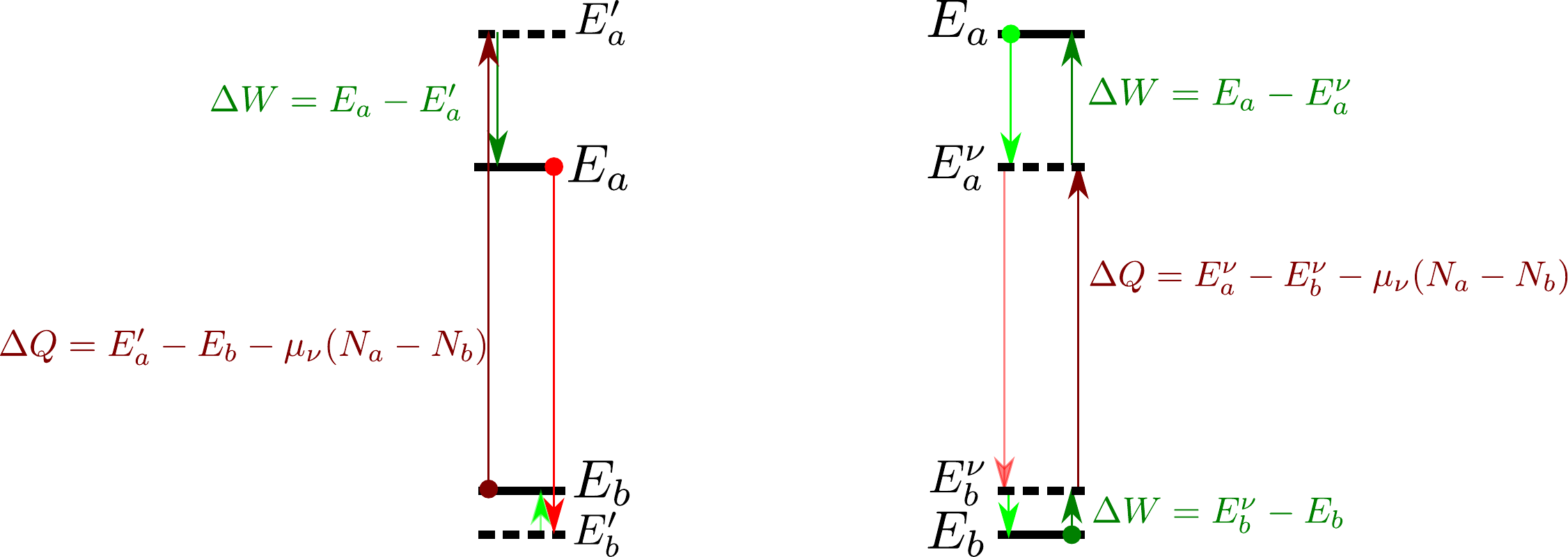}
\caption{\label{FIG:feedbackenergetics}Sketch of the instantaneous feedback scheme for the transition between states $b\to a$ (dark colors) and the reverse transition (light colors).
Mere observation of the system energy at a fixed time would yield the levels $E_a$ (solid black), but as the heat transfer (red) may not correspond to the exact energy differences, the auxiliary energies (dashed) are introduced to account for the feedback work (green). 
In the Maxwell-demon limit, we have $E_a^\nu \approx E_a$ and $E_a'\approx E_a$.
{\bf Left:} For state-conditioned feedback, the feedback work cannot depend on the reservoir $\nu$.
{\bf Right:} For jump-conditioned feedback, the feedback work is split into an energetic contribution preceding the jump and one succeeding it. 
}
\end{figure}

For example, if we assume that the feedback is conditioned on detecting the system in a particular final state $a$ with energy $E_a'$ before the control operation (see Fig.~\ref{FIG:feedbackenergetics} left panel), the feedback work cannot depend on how (e.g. by which reservoir) the system got there.
This is different from the case where the feedback is conditioned on the detection of a quantum jump from $b\to a$ due to reservoir $\nu$ with energy exchange $E_a^\nu-E_b^\nu$: As the energy exchange may depend on the reservoir $\nu$, so may the feedback work contribution, which may be further split into contributions prior and subsequent to the jump (Fig.~\ref{FIG:feedbackenergetics} right panel).
Another striking difference is that for state-conditioned feedback the heat transferred to and from the reservoir may differ for forward and backward transitions, whereas for jump-conditioned feedback it is the same (up to a sign).

To simulate merely the dynamics of the system, it is sufficient to replace the rates by their feedback-controlled versions $R_{ab}^\nu\to F_{ab}^\nu$ in~\eqref{EQ:rate},~\eqref{EQ:waiting_time}, and~\eqref{EQ:prob_jump}.
To track the dynamics of heat and entropy, it is necessary to generalize the thermodynamic laws~\eqref{EQ:first_law} and~\eqref{EQ:second_law} to instantaneous and piecewise-constant feedback:

For state-conditioned feedback, we can in analogy to~\eqref{EQ:cur_energy} write for the energy currents and the feedback work current 
\begin{align}\label{EQ:ecurfb_state}
I_E^\nu &= \sum_{ab:a\neq b} (E_a'-E_b) F_{ab}^\nu P_b\,,\qquad
I_E^{\rm fb} = \sum_\nu \sum_{ab:a\neq b} (E_a-E_a') F_{ab}^\nu P_b\,,
\end{align}
where we have used that every control operation is tied to a change in states, such that the corresponding energy difference is multiplied with the corresponding transition probability.
The total energy change (first law of thermodynamics) is then obtained by summing 
\begin{align}\label{EQ:first_law_fb}
\frac{d}{dt} \expval{E}  = \sum_\nu I_E^\nu + I_E^{\rm fb} = \sum_\nu \sum_{ab} (E_a-E_b) F_{ab}^\nu P_b\,,
\end{align}
which reflects the fact that the system is switching between energies $E_a$, see Fig.~\ref{FIG:feedbackenergetics}.
For jump-conditioned feedback, we proceed similarly any write~\eqref{EQ:cur_energy} as
\begin{align}\label{EQ:ecurfb_jump}
I_E^\nu &= \sum_{ab:a\neq b} (E_a^\nu-E_b^\nu) F_{ab}^\nu P_b\,,\qquad
I_E^{\rm fb} = \sum_\nu \sum_{ab:a\neq b} (E_b^\nu - E_b + E_a - E_a^\nu) F_{ab}^\nu P_b\,,
\end{align}
and their sum yields the same result as before. 
So far, these partitions into heat and feedback work contributions are somewhat arbitrary, they can be fixed using some microscopic intuition as we do in Secs.~\ref{SEC:4level} and~\ref{SEC:3level}.

Even when the feedback does not change the system energies (e.g., when only the transition rates are modified), the local detailed balance relation~\eqref{EQ:local_detailed_balance} will no longer hold.
To account for this and the proper corresponding energetic exchanges with the reservoirs, we parametrize the deviation from local detailed balance as~\cite{esposito2012a}
\begin{align}\label{EQ:local_detailed_balance_fb}
\frac{F_{ab}^\nu}{F_{ba}^\nu} = e^{-\beta_\nu[\Delta E_{ab}^\nu-\mu_\nu(N_a-N_b)]} e^{+\delta_{ab}^\nu}\,,
\end{align}
where $\Delta E_{ab}^\nu$ is the true heat exchange associated with the transition and $\delta_{ab}^\nu$ is a dimensionless feedback parameter~\footnote{
Note that one has the relation 
$\delta_{ab}^\nu + \delta_{ba}^\nu = \beta_\nu(\Delta E_{ab}^\nu + \Delta E_{ba}^\nu)$.}.
The feedback-free situation is thus reproduced for $\Delta E_{ab}^\nu = E_a-E_b$ and $\delta_{ab}^\nu\to 0$.

With this, the last two lines of the system entropy balance~\eqref{EQ:entropy_balance} change into
\begin{align}
\frac{d}{dt} \expval{S} &= +\sum_\nu \sum_{ab} F_{ab}^\nu P_b \ln \frac{F_{ab}^\nu P_b}{F_{ba}^\nu P_a} +\sum_\nu \sum_{ab:a\neq b} F_{ab}^\nu P_b \ln\left[ e^{\beta_\nu [\Delta E_{ab}^\nu-\mu_\nu (N_a-N_b)} e^{-\delta_{ab}^\nu}\right]\nn
&= +\sum_\nu \sum_{ab} F_{ab}^\nu P_b \ln \frac{F_{ab}^\nu P_b}{F_{ba}^\nu P_a} + \sum_\nu \beta_\nu (I_E^\nu -  \mu_\nu I_M^\nu) - \sum_\nu \sum_{ab:a\neq b} F_{ab}^\nu P_b \delta_{ab}^\nu\,.
\end{align}
As before, the first term is positive by construction and is identified with the entropy production rate of the (model) universe, the second term is still interpreted as the entropy production in the reservoirs (which are untouched by the feedback discussed here), but the last term quantifies how the feedback changes the entropy balance of the system and can thus be identified with an information current
\begin{align}\label{EQ:cur_info}
I_I \equiv +\sum_\nu \sum_{ab:a\neq b} F_{ab}^\nu P_b \delta_{ab}^\nu\,,
\end{align}
such that the second law~\eqref{EQ:second_law} generalizes to
\begin{align}\label{EQ:second_law_fb}
\sigma_\ii = \frac{d}{dt} \expval{S} - \sum_\nu \beta_\nu (I_E^\nu -  \mu_\nu I_M^\nu) + I_I  \ge 0\,.
\end{align}
Even at steady state (where $\frac{d}{dt} \expval{S} \to 0$), it is now possible to have negative entropy produced in the reservoirs ($- \sum_\nu \beta_\nu (I_E^\nu -  \mu_\nu I_M^\nu) \le 0$) as long as the information current counter-balances it such that the above modified second law~\eqref{EQ:second_law_fb} is still respected.
Of course, this treatment neglects a (possibly large) part of the entropy produced in the unspecified feedback loop, it only takes into account its effect on the system.
Maxwell demon feedbacks, in which we are mostly interested here, generate a situation where the feedback energy current is negligible $\abs{I_E^{\rm fb}} \ll \abs{I_E^\nu}$, while at the same time, the information current is substantial $\ord\{I_I\} = \ord\{\beta_\nu (I_E^\nu - \mu_\nu I_M^\nu)\}$. 
Both feedback schemes are capable of generating such scenarios.

\subsection{Example: Maxwell-demon feedback for a two-level system}

When we consider a two-level system $P_0,P_1$ with energies $E_0<E_1$ coupled to two (hot and cold) thermal reservoirs $\nu\in\{h,c\}$, the populations of its energy eigenstates follow in absence of feedback the rate equation
\begin{align}\label{EQ:rmat2level}
\frac{d}{dt} \left(\begin{array}{c}
P_0\\
P_1
\end{array}\right) 
= \left(\begin{array}{cc}
-R_{10}^c-R_{10}^h & +R_{01}^c+R_{01}^h\\
+R_{10}^c+R_{10}^h & -R_{01}^c-R_{01}^h
\end{array}\right)
\left(\begin{array}{c}
P_0\\
P_1
\end{array}\right)\,,
\end{align}
where by construction, we assume the implicit condition $T_c<T_h$ or $\beta_c > \beta_h$.
The rates satisfy local detailed balance~\eqref{EQ:local_detailed_balance}~\footnote{Often occurring rates of this form are Bose distributions $R_{10}^\nu = \Gamma_\nu [e^{\beta_\nu (E_1-E_0)}-1]^{-1}$ and $R_{01}^\nu=\Gamma_\nu\{1+[e^{\beta_\nu(E_1-E_0)}-1]^{-1}\}$ or Fermi functions
$\gamma^\nu_\upa = \Gamma_\nu [e^{\beta_\nu(E_1-E_0)}+1]^{-1}$ and $\gamma^\nu_\dna=\Gamma_\nu\{1-[e^{\beta_\nu(E_1-E_0)}+1]^{-1}\}$, where the bare rates $\Gamma_\nu$ would cancel in~\eqref{EQ:local_detailed_balance}.}.
To check thermodynamic consistency, consider the stationary energy current entering the system via the cold junction
$\bar I_E^c = (E_1-E_0) \frac{R^c_{10} R^h_{01} - R^c_{01} R^h_{10}}{R^c_{10}+R^c_{01}+R^h_{10}+R^h_{01}}
= (E_1-E_0) \frac{R^c_{10} R^h_{10} (e^{\beta_h(E_1-E_0)}-e^{\beta_c(E_1-E_0)})}{R^c_{10}+R^c_{01}+R^h_{10}+R^h_{01}}$.
This is always negative, which means that heat flows from hot to cold (Clausius formulation of the second law).
This is also confirmed by looking at the entropy production rate~\eqref{EQ:second_law}, which at steady state (where $\bar I_M^h = -\bar I_M^c$ and $\bar I_E^h = -\bar I_E^c$ and the system contribution is negligible) simplifies to
$\dot{\bar\sigma}_\ii = (\beta_h-\beta_c) \bar I_E^c \ge 0$.

In presence of feedback, we now consider an increase or a decrease (compared to the feedback-free case) of the bare tunnelling rate to reservoir $\nu$ parametrized by feedback parameters $\delta^\nu_0$ and $\delta^\nu_1$, such that the feedback-controlled rates relate to the feedback-free rates as
\begin{align}
F_{10}^c &= R_{10}^c e^{+\delta^c_0}\,,\qquad
F_{01}^c = R_{01}^c e^{+\delta^c_1}\,,\nn
F_{10}^h &= R_{10}^h e^{+\delta^h_0}\,,\qquad
F_{01}^h = R_{01}^h e^{+\delta^h_1}\,,
\end{align}
and we assume the energies to remain unperturbed $\Delta E_{ab}^\nu = E_a-E_b$, such that $\delta_{01}^\nu = \delta^\nu_1-\delta^\nu_0 = -\delta_{10}^\nu$.
This can significantly alter the transport characteristics~\footnote{Consider the extreme limit $\delta^c_0 \to +\infty$, $\delta^c_1 \to -\infty$, $\delta^h_0\to-\infty$, and $\delta^h_1\to +\infty$, where the two-level system couples only to the cold reservoir when it is in its ground state and to the hot reservoir while it is in its excited state.
In consequence, the energy can only flow from cold to hot reservoir.}.
Accordingly, we just replace $R_{ij}^\nu \to F_{ij}^\nu$ in the rate matrix~\eqref{EQ:rmat2level}.
Then, the stationary energy current entering the system from the cold junction becomes (for Maxwell-demon type feedbacks we have by energy conservation $\bar I_E^h = -\bar I_E^c$)
\begin{align}
\bar I_E^c &=(E_1-E_0) \frac{F^c_{10} F^h_{01} - F^c_{01} F^h_{10}}{F^c_{10}+F^c_{01}+F^h_{10}+F^h_{01}}\nn
&= (E_1-E_0) \frac{R^c_{10} R^h_{10} (e^{\beta_h(E_1-E_0)} e^{\delta^c_0+\delta^h_1} - e^{\beta_c(E_1-E_0)} e^{\delta^c_1+\delta^h_0})}{F^c_{10}+F^c_{01}+F^h_{10}+F^h_{01}}\,,
\end{align}
which is positive if and only if $\delta^c_0+\delta^h_1-\delta^c_1-\delta^h_0 = \delta_{01}^h-\delta_{01}^c > (\beta_c-\beta_h)(E_1-E_0)$.
Thus, with feedback we can direct the stationary energy current from the cold reservoir to the hot one by using sufficiently strong feedback parameters $\delta^\nu_a$.
The first law reads as before (Maxwell demon feedback by assumption does not inject energy), but the information current modifies the second law, such that the entropy production rate under feedback~\eqref{EQ:second_law_fb} becomes manifestly positive
\begin{align*}
\dot{\bar\sigma}_\ii = (\beta_h-\beta_c) \bar I_E^c + \bar I_I \ge 0\qquad:\qquad
\bar I_I = \frac{\delta^c_0-\delta^c_1-\delta^h_0+\delta^h_1}{E_1-E_0} \bar I_E^c = \frac{\delta_{01}^h-\delta_{01}^c}{E_1-E_0} \bar I_E^c\,.
\end{align*}

\section{Autonomous systems}

\begin{figure}
\begin{tabular}{ccc}
\includegraphics[width=0.4\textwidth]{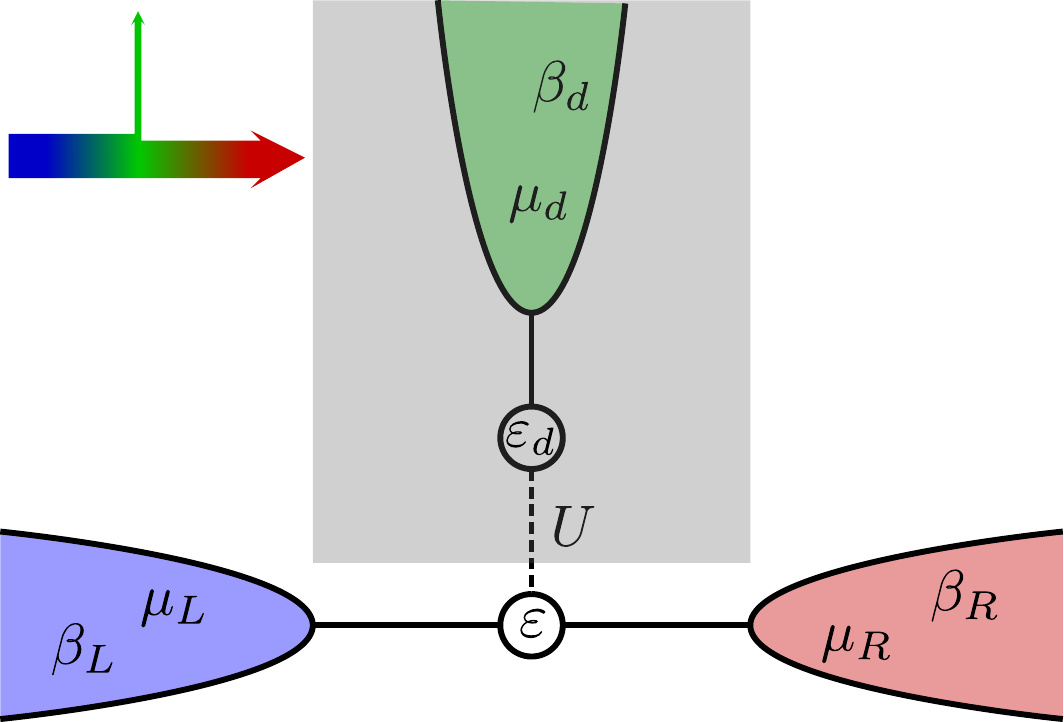} &
\hspace{2cm} &
\includegraphics[width=0.4\textwidth]{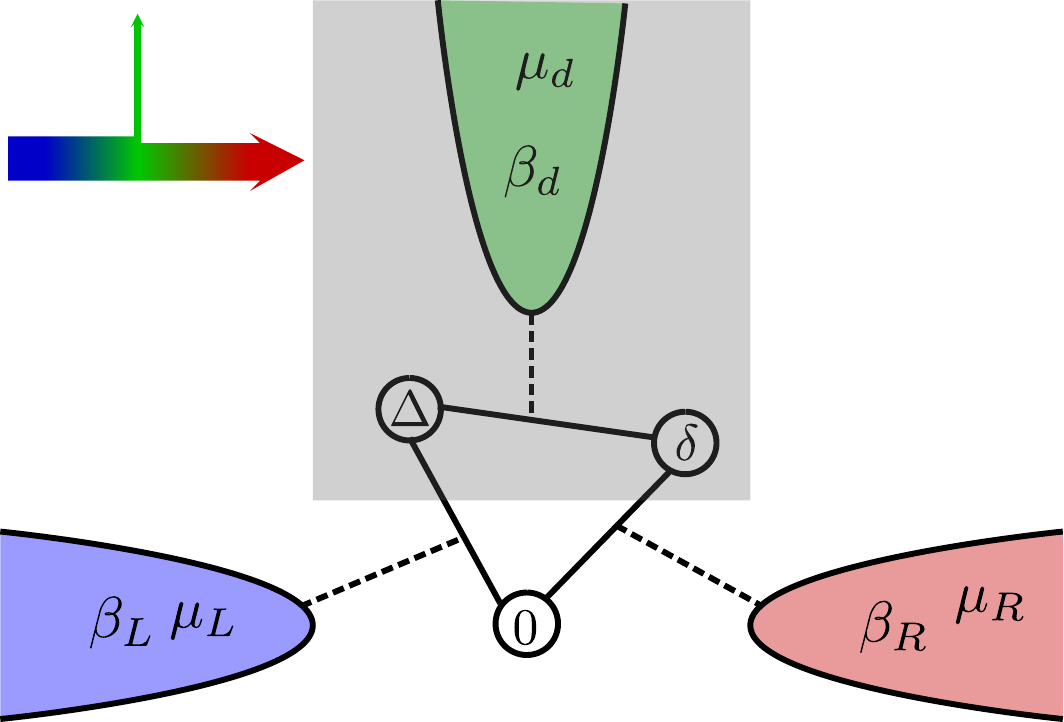}
\end{tabular}
\caption{\label{FIG:sketch_nlevelsystem}
Autonomous demonic refrigerators considered in the subsections below. Solid lines denote tunnel junctions whereas dashed lines indicate merely energetic exchanges. With three (blue left, red right, and green demon) terminals at inverse temperatures $\beta_d > \beta_L > \beta_R$ (chemical potentials can be chosen to vanish throughout $\mu_\nu=0$), it is possible to cool the left (blue) reservoir.
In specific regimes, the devices act like autonomous Maxwell demons in the sense that in ignorance of the demon reservoir (shaded region), heat flows from cold (blue) to hot (red) with negligible energy loss (green) across the demon junction (arrows on top left).
{\bf Left}: The implementation with two capacitively interacting quantum dots (circles) with variable particle number leads to a bipartite system (bottom) - controller (shaded) setup with four levels.
{\bf Right}: An analog non-bipartite three-level system can be implemented by the singly-charged sector of three quantum dots with tunnelling events between them triggered by the reservoirs.
}
\end{figure}

The implementation of external feedback control loops is experimentally challenging~\cite{chida2017a} and has shortcomings in its theoretical discussion (effective thermodynamic discussion, highly idealized assumptions on measurement and feedback speed, \ldots).
On-chip implementations of feedback loops in higher-dimensional rate equations do not suffer from these problems but allow for convenient cycle analyses~\cite{mayrhofer2021a}.
Indeed, ideal Maxwell demons can -- at least within the limits of perturbative descriptions -- be very closely approximated~\cite{sanchez2019b}.
Aiming at smallest systems, we focus here on setups that can direct the flow of energy from a cold reservoir towards a hot one by sacrificing a small fraction of the energy to a ultra-cold third terminal, see Fig.~\ref{FIG:sketch_nlevelsystem}.

\subsection{Autonomous four-level system: Interacting quantum dots}\label{SEC:4level}

We first review a bipartite model shown in Fig.~\ref{FIG:sketch_nlevelsystem} left panel, that has been discussed theoretically with focus on battery charging~\cite{strasberg2013a} and also has been implemented experimentally~\cite{koski2015a}.
It consists of two interacting two-level subsystems (quantum dots), such that our total system has four levels with the corresponding occupation probabilities
$P_{00},P_{10},P_{01},P_{11}$ of finding the first system in state $0,1,0,1$ while the second is in states $0,0,1,1$, respectively.
The first system is tunnel-coupled to two (left and right) reservoirs with additive transition rates, whereas the second is only coupled to one reservoir.
Due to Coulomb interaction, the transition rates of one subsystem may depend on the state of the other.
The rate matrix ${\cal R}$ becomes
\begin{align*}
\left(\begin{array}{cccc}
-\gamma^d_\upa-\gamma^L_\upa-\gamma^R_\upa & \gamma^L_\dna+\gamma^R_\dna & \gamma^d_\dna & 0\\
\gamma^L_\upa+\gamma^R_\upa & -\gamma^{dU}_\upa - \gamma^L_\dna-\gamma^R_\dna & 0 & \gamma^{dU}_\dna\\
\gamma^d_\upa & 0 & -\gamma^d_\dna-\gamma^{LU}_\upa-\gamma^{RU}_\upa & \gamma^{LU}_\dna+\gamma^{RU}_\dna\\
0 & \gamma^{dU}_\upa & \gamma^{LU}_\upa+\gamma^{RU}_\upa & -\gamma^{dU}_\dna-\gamma^{LU}_\dna-\gamma^{RU}_\dna
\end{array}\right)\,.
\end{align*}
The rates can be written (beyond the wideband approximation) as products of bare tunneling rates $\Gamma_\nu^{(U)}$ and Fermi functions $f_\nu^{(U)}$
\begin{align}
\gamma^\nu_\upa &= \Gamma_\nu f_\nu\,,\qquad
\gamma^\nu_\dna = \Gamma_\nu (1-f_\nu)\,,\nn
\gamma^{\nu U}_\upa &= \Gamma_\nu^U f_\nu^U\,,\qquad
\gamma^{\nu U}_\dna = \Gamma_\nu^U (1-f_\nu^U)\,,
\end{align}
where the left and right Fermi functions are given by $f_\nu = [e^{\beta_\nu(\epsilon-\mu_\nu)}+1]^{-1}$ and $f_\nu^U = [e^{\beta_\nu(\epsilon+U-\mu_\nu)}+1]^{-1}$ for the system reservoirs $\nu \in \{L,R\}$ and analogously $f_d=[e^{\beta_d(\epsilon_d-\mu_d)}+1]^{-1}$ and $f_d^U = [e^{\beta_d(\epsilon_d+U-\mu_d)}+1]^{-1}$ for the demon reservoir.

\subsubsection{Function: Cooling the left reservoir}

\begin{figure}
\includegraphics[width=0.8\textwidth]{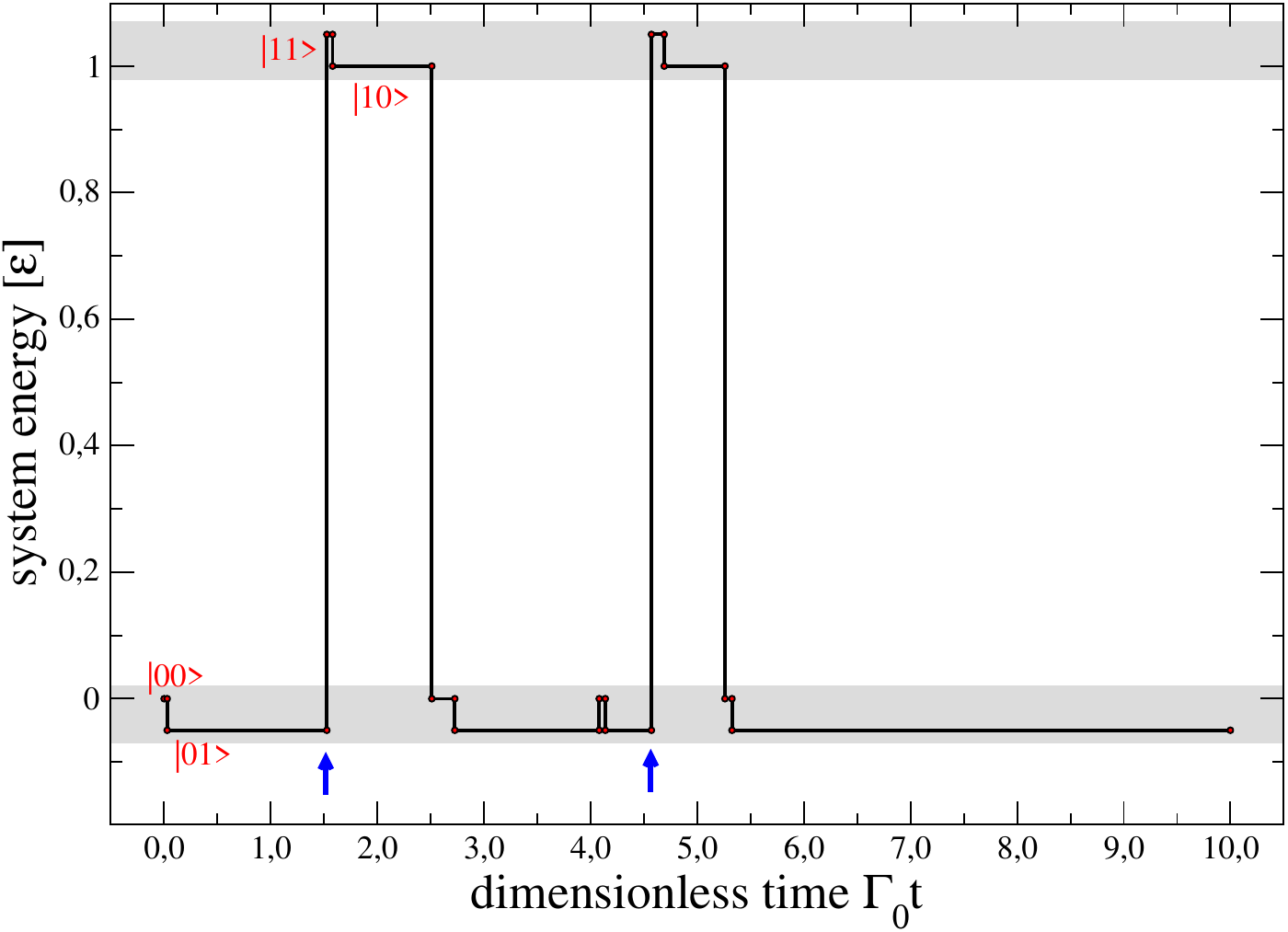}
\caption{\label{FIG:4leveltrj}Typical trajectory (see Sec.~\ref{SEC:stochastic_propagation}) for the four-level system in the operational regime of cooling functionality for the left reservoir.
The system is initialized in state $|00\rangle$, and right afterwards the controller dot is loaded (changing its energy to $\epsilon_d$ chosen negative here) leading to state $|01\rangle$.
Then, occasionally the dot is loaded from the left (the right tunneling rate at this energy is suppressed) reservoir (transferring energy $\epsilon+U$ into the system, cooling functionality).
However, for a successful cycle (blue arrows), the controller dot unloads quickly to reach state $|10\rangle$ (dissipating heat $\epsilon_d+U$ into the demon reservoir), the system can then unload an electron with energy $\epsilon$ to the right (the left tunnelling rate at this energy is suppressed), effectively transferring energy $\epsilon$ to the right reservoir.
The cycle is then closed by the controller dot loading again leading to state $|01\rangle$.
Shaded regions mark the coarse-grained states of the system dot~\eqref{EQ:cg1}.
Other parameters: $\Gamma_L^U = \Gamma_R = \Gamma_0$, $\Gamma_d = \Gamma_d^U = 10 \Gamma_0$, $\Gamma_L = \Gamma^R_U = 0$, $\epsilon_d = -.05 \epsilon$, $U=0.1\epsilon$, $\mu_\nu=0$, $\beta_d \epsilon = 100$, $\beta_R \epsilon = 1$, $\beta_L \epsilon = 2$.
}
\end{figure}

To implement cooling on the left reservoir while $\beta_L>\beta_R$ and $\mu_L=\mu_R$, one needs a cold demon reservoir and energies and chemical potentials tuned such that $f_d \to 1$ and $f_d^U \to 0$, and furthermore some asymmetry in the left and right tunnelling rates, such that $\Gamma_L^U, \Gamma_R \gg \Gamma_L,\Gamma_R^U$.
A typical trajectory in this operational regime is depicted in Fig.~\ref{FIG:4leveltrj}.
Since the device operates stochastically, one has unsuccessful loops like $\ket{01}\to\ket{00}\to\ket{01}$ (which has zero net energy exchange), but in the chosen regime, successful loops like $\ket{01} \to \ket{11} \to \ket{10} \to \ket{00} \to \ket{01}$ become likely.
This loop effectively transfers energy $\epsilon+U$ out of the left reservoir to the demon reservoir with fraction $U$ and the right reservoir with fraction $\epsilon$.
Thus, the regime of a proper Maxwell demon is reached when we have $U \ll \epsilon$ while the directionality of all transitions is maintained.
The required conditions on parameters $\Gamma_d,\Gamma_d^U \gg \Gamma_{L/R},\Gamma_{L/R}^U$ and  $\Gamma_L^U, \Gamma_R \gg \Gamma_L, \Gamma_R^U$ and $\beta_d U \gg 1$ can be simultaneously fulfilled~\cite{strasberg2013a}.

\subsubsection{Relation to the feedback-controlled model}

We can introduce the marginal probabilities for the state of the controlled system by summing over the states of the demon dot
\begin{align}\label{EQ:cg1}
P_0 = P_{00}+P_{01}\,,\qquad
P_1 = P_{10}+P_{11}\,,
\end{align}
and when constructing the differential equations for them we rewrite $P_{ij}=\frac{P_{ij}}{P_i} P_i \approx (\lim\limits_{\Gamma_d,\Gamma_d^U\to\infty}\frac{P_{ij}}{P_i}) P_i$.
The in reality time-dependent prefactor can in the limit of a very fast demon reservoir be replaced by a constant
\begin{align}
\lim_{\Gamma_d\to\infty} \frac{P_{00}}{P_0} &= \frac{\gamma^d_\dna}{\gamma^d_\dna+\gamma^d_\upa} = 1-f_d\,,\qquad
\lim_{\Gamma_d\to\infty} \frac{P_{01}}{P_0} =  \frac{\gamma^d_\upa}{\gamma^d_\dna+\gamma^d_\upa} = f_d\,,\\
\lim_{\Gamma_d^U\to\infty} \frac{P_{10}}{P_1} &= \frac{\gamma^{dU}_\dna}{\gamma^{dU}_\dna+\gamma^{dU}_\upa} = 1-f_d^U\,,\qquad
\lim_{\Gamma_d^U\to\infty} \frac{P_{11}}{P_1} =  \frac{\gamma^{dU}_\upa}{\gamma^{dU}_\dna+\gamma^{dU}_\upa} = f_d^U\,,\nonumber
\end{align}
where $0 \le f_d \le 1$ and $0 \le f_d^U \le 1$,
to obtain a coarse-grained rate equation~\cite{esposito2012b,ehrich2021a,strasberg2022} for the probabilities $P_0,P_1$ of the system dot being empty or filled like shown by the shaded regions of Fig.~\ref{FIG:4leveltrj}, respectively. 
In the limit where $f_d\to 1$ and $f_d^U \to 0$, the coarse-grained rate matrix simplifies~\footnote{
This happens at low demon reservoir temperatures $\beta_d U \gg 1$ with e.g. $\epsilon_d \approx \mu_d-U/2$.
It is also the limit where the measurement becomes perfect and system and demon dot are always anti-correlated.}
further into
\begin{align}
{\cal R}_{\rm cg} = \left(\begin{array}{cc}
-(\gamma^{LU}_\upa+\gamma^{RU}_\upa) & (\gamma^L_\dna+\gamma^R_\dna)\\
+ (\gamma^{LU}_\upa+\gamma^{RU}_\upa) & -(\gamma^L_\dna+\gamma^R_\dna)
\end{array}\right)\,.
\end{align}
In this, we see that the effective transition energy for the relaxation rate is different from the transition energy for the excitation, such that the autonomous four-level system effectively implements a state-conditioned feedback discussed in Sec.~\ref{SEC:feedback_control}.
The relevant ratios of effective transition rates for our entropic balance can be written as
\begin{align}
\frac{F_{01}^\nu}{F_{10}^\nu} &\to \frac{\gamma^\nu_\dna}{\gamma^{\nu U}_\uparrow} = \frac{\gamma^\nu_\dna}{\gamma^{\nu}_\uparrow} \frac{\gamma^\nu_\upa}{\gamma^{\nu U}_\uparrow}
= e^{\beta_\nu (\epsilon-\mu_\nu)} \frac{\gamma^\nu_\upa}{\gamma^{\nu U}_\uparrow}
\,,\nn
\frac{F_{10}^\nu}{F_{01}^\nu} &\to \frac{\gamma^{\nu U}_\uparrow}{\gamma^\nu_\dna} = \frac{\gamma^{\nu U}_\upa}{\gamma^{\nu U}_\dna} \frac{\gamma^{\nu U}_\dna}{\gamma^\nu_\dna}
= e^{-\beta_\nu(\epsilon+U-\mu_\nu)} \frac{\gamma^{\nu U}_\dna}{\gamma^\nu_\dna}\,,
\end{align}
where we have used that the heat transfer for an initially filled system dot (and empty demon dot) is $\epsilon$, whereas it is $\epsilon+U$ for an initially empty system (and filled demon dot).
From this, we can identify with~\eqref{EQ:local_detailed_balance_fb} the energy differences $\Delta E_{01}^\nu = -\epsilon$ and $\Delta E_{10}^\nu = \epsilon+U$, which can be satisfied~\footnote{
These shifts can be understood as unloading of the dot transfers energy $\epsilon$ to the left or right lead, wheras loading requires energy $\epsilon+U$ (the difference is transferred to the demon reservoir as effective feedback work).}
with energies $E_0=0=E_0'$, and $E_1=\epsilon$ and $E_1' = \epsilon+U$.
The feedback parameters become
\begin{align}
\delta^\nu_{01} = \ln \frac{\gamma^\nu_\upa}{\gamma^{\nu U}_\uparrow}\,,\qquad
\delta^\nu_{10} = \ln \frac{\gamma^{\nu U}_\dna}{\gamma^\nu_\dna}\,,
\end{align}
such that $\delta^\nu_{01}+\delta^\nu_{10} = \beta_\nu U$.

We can thus write for the coarse-grained energy currents
\begin{align}
I_{E, \rm cg}^\nu = (\epsilon+U) F_{10}^\nu P_0 - \epsilon F_{01}^\nu P_1\,,\qquad
I_E^{\rm fb} = -U (F_{01}^L + F_{01}^R) P_1\,,
\end{align}
These currents match what is obtained from the microscopic four-level model in the limit of a fast and precise demon.

Accordingly, the information current becomes
\begin{align}
I_I &= \sum_\nu \left[F_{01}^\nu P_1 \ln \frac{\gamma^\nu_\upa}{\gamma^{\nu U}_\upa} + F_{10}^\nu P_0 \ln \frac{\gamma^{\nu U}_\dna}{\gamma^\nu_\dna}\right]\,.
\end{align}
As the coarse-grained entropy production $\dot{\bar\sigma}_{\ii,{\rm cg}} = -\beta_L \bar I_{E, \rm cg}^L - \beta_R \bar I_{E, \rm cg}^R + \bar I_I \le \dot{\bar\sigma}_\ii$ under-estimates the full entropy production~\cite{esposito2012b} $\dot{\bar\sigma}_\ii =  -\beta_L \bar I_E^L - \beta_R \bar I_E^R - \beta_d \bar I_E^d$, and the coarse-grained energy currents converge to the ones from the microscopic model $\bar I_{E, \rm cg}^\nu \to \bar I_E^\nu$, it follows that the information current is smaller than the entropy production in the controller reservoir, such that the information gained in the measurement is only partially used in the feedback loop. 
Furthermore, it can be concluded that when we evaluate the thermodynamic uncertainty relation~\cite{barato2015a} $\dot{\bar\sigma}_\ii \frac{\bar S}{\bar I^2} \ge 2$, (where $\bar I$ and $\bar S$ are energy or matter current and its associated noise over an arbitrary junction), with the coarse-grained entropy production instead, we can find regimes where it is violated.

\subsection{Autonomous three-level system}\label{SEC:3level}

We now consider a three-level system as in Fig.~\ref{FIG:sketch_nlevelsystem} right panel, with its three transitions selectively driven by three independent reservoirs.
Such a system could be the single-charged sector of a triple quantum dot (see e.g. Ref.~\cite{seo2013a} for an experimental implementation).
We assume that the left (cold) reservoir drives the large transition between ground and second excited state, the right (hot) reservoir the smaller transition between ground and first excited state, and the demon reservoir (coldest) the remaining one between first and second excited state~\footnote{
This is different from the commonly used quantum absorbtion refrigerator (QAR)~\cite{levy2012b,linden2010a}:
First, in a QAR the hot reservoir drives the large excitation above the ground state.
Second, a QAR would require the transition between the two excited states driven by a work reservoir (e.g. a reservoir of substantially higher temperature than the hot and cold ones).
Third, the operational regime of a QAR is characterized by a non-negligible energy current from the work reservoir $\bar I_E^w = \bar I_E^L/\kappa \ge \bar I_E^L/\kappa_{\rm Ca}$, with 
the cofficient of performance $\kappa \le \kappa_{\rm Ca}$ limited by its Carnot value $\kappa_{\rm Ca} = \beta_R/(\beta_L-\beta_R)$.
Unlike a Maxwell demon, a QAR is thus energy-dominated, it could be implemented by reverting the temperature hierarchy to $\beta_d < \beta_L < \beta_R$.}.
We consider here the limit of a low-temperature demon reservoir with negligible energetic contribution that reverses the usual heat flow.
In the basis $P_0, P_\delta, P_\Delta$, the rate matrix for our model reads
\begin{align}
{\cal R} = 
\left(\begin{array}{ccc}
-\gamma^L_\upa - \gamma^R_\upa & \gamma^R_\dna & \gamma^L_\dna\\
\gamma^R_\upa & -\gamma^R_\dna-\gamma^d_\upa & \gamma^d_\dna\\
\gamma^L_\upa & \gamma^d_\upa & -\gamma^L_\dna-\gamma^d_\dna
\end{array}\right)\,,
\end{align}
where $\gamma^\nu_\upa$ denotes the excitation and $\gamma^\nu_\dna$ the de-excitation rate due to reservoir $\nu$.

\subsubsection{Function: Cooling the left reservoir}

\begin{figure}
\includegraphics[width=0.8\textwidth]{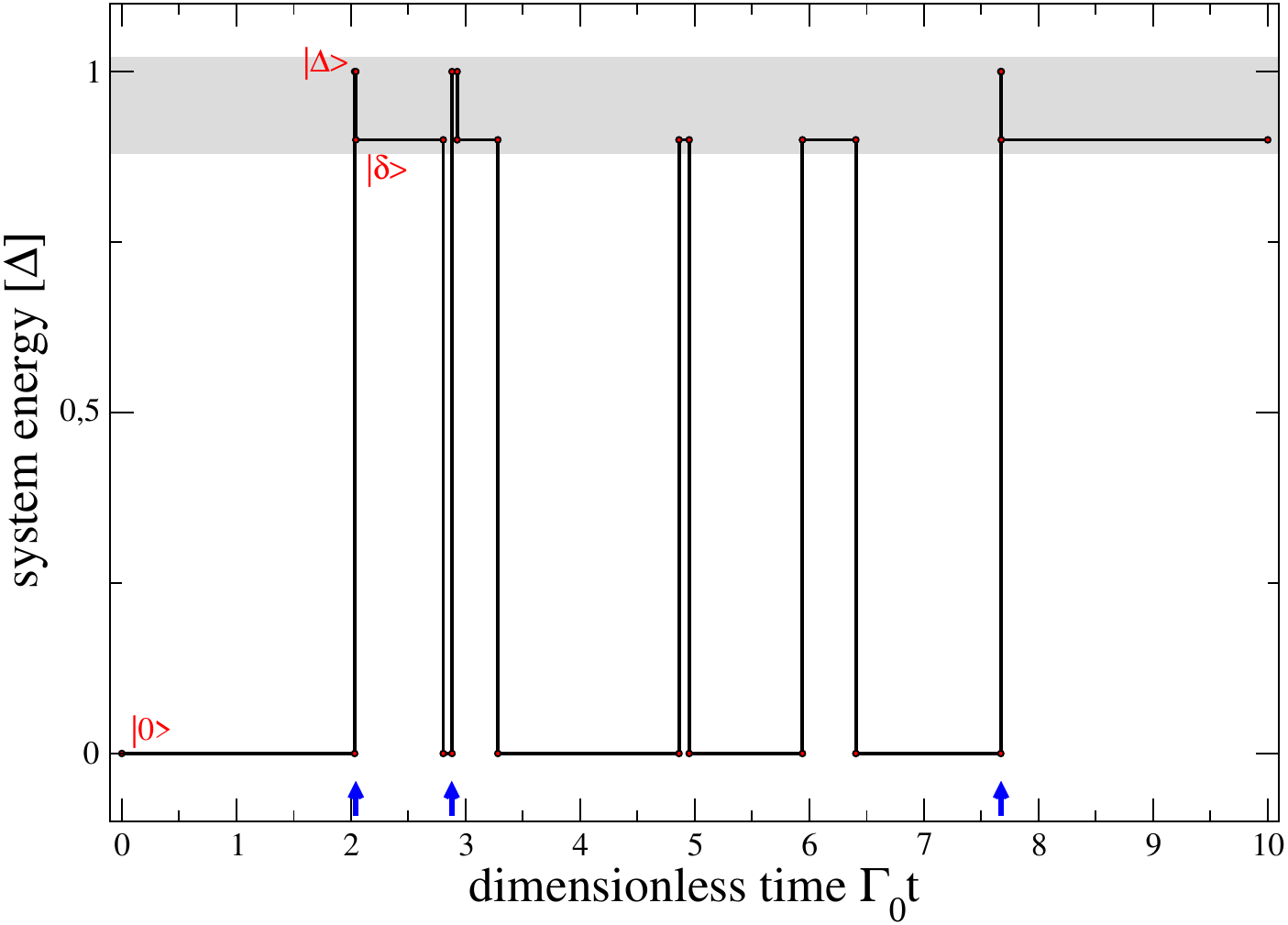}
\caption{\label{FIG:3leveltrj}Typical trajectory (see Sec.~\ref{SEC:stochastic_propagation}) for the three-level system in the operational regime of cooling functionality for the left reservoir.
The system is initialized in state $|0\rangle$, and is occasionally excited to $|\delta\rangle$ by the right, or to $|\Delta\rangle$ by the left reservoir.
In the first case, the excitation is mostly given back to the right reservoir as the demon reservoir has a much lower temperature that prohibits further excitation (e.g. middle of the plot).
From state $|\Delta\rangle$ though (blue arrows), the demon reservoir quickly induces a relaxation to state $\delta$, from where via the transition to $|0\rangle$ most of the heat is then dissipated into the right reservoir (thereby closing the cooling cycle).
The shaded region marks the coarse-grained states~\eqref{EQ:cg2}.
Other parameters: $\gamma^\nu_\upa = \Gamma_\nu n_\nu$ and $\gamma^\nu_\dna = \Gamma_\nu (1+n_\nu)$ with Bose distributions $n_L =[e^{\beta_L\Delta}-1]^{-1}$, $n_R =[e^{\beta_R\delta}-1]^{-1}$, $n_d =[e^{\beta_d(\Delta-\delta)}-1]^{-1}$ and $\Gamma_L = \Gamma_R = \Gamma_0$, $\Gamma_d = 100 \Gamma_0$, $\delta = 0.9\Delta$, $\beta_d = 100 \beta_R$, $\beta_L = 2 \beta_R$, $\beta_R \Delta = 1$.
}
\end{figure}

In the desired operational mode, the second excited state is occasionally reached from the ground state by the absorbtion of energy $\Delta$ from the cold left reservoir (implementing cooling).
The ultra-cold demon reservoir then takes a fraction $\Delta-\delta$ of that energy to transfer the system from the second to its first excited state.
Finally, the hot reservoir absorbs heat $\delta$ by allowing the transfer from the first excited state back to the ground state, closing the cycle.
The opposite process is blocked by the ultra-cold demon reservoir in the intended regime, see Fig.~\ref{FIG:3leveltrj} for a typical trajectory.
Thus, for a successful cycle $\ket{0}\to \ket{\Delta}\to \ket{\delta}\to \ket{0}$, energy $\Delta$ is taken from the left reservoir, while energies $\Delta-\delta$ and $\delta$ are dissipated into the demon and right reservoirs, respectively. 
The Maxwell demon limit is thus approached when $\delta \to \Delta$, while the directionality of transitions is maintained.
For our model, this requires $\Gamma^d \gg \Gamma^L,\Gamma^R$ and $\beta_d (\Delta-\delta) \gg 1$.

The stationary currents entering the system from the reservoirs read
\begin{align}
\bar I_E^L &= \frac{\Delta(\gamma^L_\upa \gamma^d_\dna \gamma^R_\dna- \gamma^L_\dna \gamma^d_\upa \gamma^R_\upa)}
{[\gamma^L \gamma^d-\gamma^L_\dna \gamma^d_\dna]+
[\gamma^L \gamma^R-\gamma^L_\upa \gamma^R_\upa]+
[\gamma^R \gamma^d-\gamma^R_\dna \gamma^d_\upa]}\,,
\end{align}
where $\gamma^\nu \equiv \gamma^\nu_\upa + \gamma^\nu_\dna$.
The other currents are tightly coupled $\bar I_E^R = -\frac{\delta}{\Delta} \bar I_E^L$ and $\bar I_E^d = -\frac{\Delta-\delta}{\Delta} \bar I_E^L$,  
such that the sum of all stationary currents vanishes (first law) and by using $\gamma^d_\upa \to 0$ we can implement cooling of the cold left reservoir ($\bar I_E^L > 0$).

At steady-state, the entropy production rate~\eqref{EQ:second_law} thus reads
\begin{align}
\dot{\bar\sigma}_\ii = -\beta_L \bar I_E^L - \beta_R \bar I_E^R - \beta_d \bar I_E^d = \left(\beta_R \frac{\delta}{\Delta} + \beta_d \frac{\Delta-\delta}{\Delta} - \beta_L\right)\bar I_E^L \ge 0\,.
\end{align}
Thus, cooling occurrs when the prefactor in round brackets is positive, i.e., when $\beta_d-\beta_L \ge (\beta_d-\beta_R) \delta/\Delta$.
Furthermore, by choosing $\Delta-\delta\to 0$ the energy dissipated into the demon reservoir can be made arbitrarily small while the entropy produced there remains large due to its low temperature $\beta_d (\Delta-\delta) \gg 1$.
This already indicates a demon operational mode~\footnote{
An efficiency of the device could be defined by
$\eta = -\bar I_E^R/\bar I_E^L \Theta(\bar I_E^L) = \delta/\Delta \Theta(\bar I_E^L) \le 1 - (\beta_L-\beta_R)/(\beta_d-\beta_R)$,
which is upper-bounded by unity as $\beta_d\to\infty$.}, but we can make the relation to the external control loop more explicit by applying a similar coarse-graining procedure as before.

\subsubsection{Relation to the feedback-controlled model}

In the limit where the demon reservoir rates are both significantly larger than the others while maintaining local detailed balance~\eqref{EQ:local_detailed_balance} (consider e.g. $\gamma^d_\dna = \Gamma_d (1+n_d)$, $\gamma^d_\upa = \Gamma_d n_d$ with Bose distribution $n_d = [e^{\beta_d (\Delta-\delta)}-1]^{-1}$ in the limit $\Gamma_d\to\infty$), we can formally coarse-grain (a similar argument is used in Sec.~V of Ref.~\cite{strasberg2014a}) the rate equation by introducing the probability to be in any of the excited states~\footnote{Technically, we construct the equation for $\dot P_{1}$ from the rate equation and rewrite $P_\delta = (P_\delta/P_1) P_1 \to [\gamma^d_\dna/(\gamma^d_\dna+\gamma^d_\upa)] P_1$ and $P_\Delta = (P_\Delta/P_1) P_1 \to [\gamma^d_\upa/(\gamma^d_\dna+\gamma^d_\upa)] P_1$ in all equations.} as indicated by the shaded region in Fig.~\ref{FIG:3leveltrj}
\begin{align}\label{EQ:cg2}
P_{1} = P_\delta+P_\Delta\,.
\end{align}
Then, the coarse-graining approximation can be implemented by using that the conditional probabilities of being in one of the excited states are given by
\begin{align*}
\lim_{\gamma^d_{\upa/\dna}\to\infty} \frac{P_\delta}{P_{1}} &= \frac{\gamma^d_\dna}{\gamma^d_\dna+\gamma^d_\upa} 
= \frac{1}{1+e^{-\beta_d(\Delta-\delta)}}\,,\qquad
\lim_{\gamma^d_{\upa/\dna}\to\infty} \frac{P_\Delta}{P_{1}} = \frac{\gamma^d_\upa}{\gamma^d_\dna+\gamma^d_\upa} = \frac{1}{1+e^{+\beta_d(\Delta-\delta)}}\,,
\end{align*}
such that the coarse-grained rate matrix for the states $(P_0,P_1)$ becomes
\begin{align}
{\cal R}_{\rm cg} = 
\left(\begin{array}{cc}
-\gamma^L_\upa-\gamma^R_{\upa} & +\gamma^L_\dna[1+e^{+\beta_d(\Delta-\delta)}]^{-1}+\gamma^R_\dna[1+e^{-\beta_d(\Delta-\delta)}]^{-1}\\
+\gamma^L_\upa+\gamma^R_{\upa} & -\gamma^L_\dna[1+e^{+\beta_d(\Delta-\delta)}]^{-1}-\gamma^R_\dna[1+e^{-\beta_d(\Delta-\delta)}]^{-1}
\end{array}\right)\,.
\end{align}
Since the ratios of the $\gamma^\nu_\dna$ and $\gamma^\nu_\upa$ already give the proper heat exchanges with those reservoirs, the three-level model thus effectively implements jump-conditioned feedback from Sec.~\ref{SEC:feedback_control}.
From~\eqref{EQ:local_detailed_balance_fb} we can directly identify the feedback parameters as
\begin{align}
\delta_{01}^L = \ln \frac{1}{1+e^{+\beta_d(\Delta-\delta)}} = -\delta_{10}^L\,,\qquad
\delta_{01}^R = \ln \frac{1}{1+e^{-\beta_d(\Delta-\delta)}} = -\delta_{10}^R\,,
\end{align}
such that $\delta_{01}^\nu + \delta_{10}^\nu = 0$.
In the zero-temperature demon limit, the cold reservoir is thus associated with divergent (infinitely strong) feedback. 

The stationary energy current from the left reservoir becomes 
\begin{align}
\bar I_{E,\rm cg}^L = -\Delta F_{01}^L \bar P_{12}+\Delta F_{10}^L \bar P_0 = 
\Delta \frac{\gamma^L_\upa \gamma^R_\dna - e^{-\beta_d (\Delta-\delta)} \gamma^L_\dna \gamma^R_\upa}{\gamma^L_\upa + \gamma^R_\upa + \gamma^R_\dna + e^{-\beta_d(\Delta-\delta)} (\gamma^L_\dna + \gamma^L_\upa + \gamma^R_\upa)}\,,
\end{align}
such that cooling is possible under the same conditions as before.
The energy current from the hot reservoir is fixed by the tight coupling condition $\bar I_{E, \rm cg}^R = -\frac{\delta}{\Delta} \bar I_{E, \rm cg}^L$ and also the energy current injected by the feedback becomes $\bar I_E^{\rm fb} = -\frac{\Delta-\delta}{\Delta} \bar I_{E, \rm cg}^L$ (as one may verify).
These currents match what is obtained for the microscopic three-level model in the limit $\Gamma_d\to\infty$ for the left, right, and controller reservoir, respectively.

Accordingly, the entropy produced in the demon reservoir is given by $\dot{\bar\sigma}_\ii^d = \beta_d \frac{\Delta-\delta}{\Delta} \bar I_E^L$.
This matches the effective information current resulting from the feedback, which becomes
\begin{align}
\bar I_I = \sum_\nu \delta_{01}^\nu (F_{01}^\nu \bar P_1 - F_{10}^\nu \bar P_0) = \frac{\beta_d (\Delta-\delta)}{\Delta} \bar I_E^L\,.
\end{align}
The three-level implementation is thus more efficient than the four-level version not only by its reduced complexity, but also its more efficient use of information: The information current due to the demon precisely corresponds to the entropy produced in the feedback loop.
The drawback is that there is no parameter to turn off the feedback smoothly (like $U$ was used in the four-level model).

\subsubsection{Electronic implementation}

One possibility to implement an electronic three-level system is in the Coulomb-blockade regime of two Coulomb-coupled quantum dots with effective single-particle onsite energies $\Delta$ and $\delta$.
Preferentially, the left (cold) reservoir should load and unload a dot mode with charging energy $\Delta$, the right (hot) reservoir should load and unload a mode with charging energy $\delta$. 
The selectivity of the coupling may be realized by using peaked reservoir spectral functions (which may be generated or enhanced further by connecting over additional fine-tuned quantum dots~\cite{ehrlich2021a}) resonant with the corresponding transition energies $\Delta$ and $\delta$, respectively.
The remaining transition between the two singly-charged modes is made possible by the action of a third (demon) reservoir, which leaves the particle number in the system constant~\footnote{
Any capacitive interaction with coupling operator $S=u_1 d_1^\dagger d_1 + u_2 d_2^\dagger d_2$ will trigger transitions between eigenmodes of the system $S = \alpha_\delta c_\delta^\dagger c_\delta + \alpha_\Delta c_\Delta^\dagger c_\Delta + [\tau c_\Delta^\dagger c_\delta + {\rm h.c.}]$.}.

If one aims at setups where one has only energy but no particle exchange, one would need three particle-conserving reservoirs.
Then, an electronic implementation of a three-level system would be possible using the singly charged sector of three tunnel-coupled quantum dots $H_S^1 = \sum_{ij} \tau_{ij} d_i^\dagger d_j$ coupled to three particle-conserving reservoirs.
An original interaction coupling to $S_\nu \propto d_i^\dagger d_i$ will then generate transition operators and energy renormalizations for all triple dot modes. 
Also here, the selectivity of the interaction could be implemented by tuning the reservoir parameters accordingly. 

The rates for the demon reservoir in the double-dot case or all reservoirs in the singly-charged sector of the triple dot case can be microscopically calculated~\cite{schaller2014}.
To see that, consider a fermionic reservoir $H_B = \sum_k \epsilon_k c_k^\dagger c_k$ with a coupling of the form
\begin{align}
H_I = S \otimes \left[\sum_{kq} t_{kq} c_k^\dagger c_q - \sum_k t_{kk} f(\epsilon_k)\right]\equiv S \otimes B\,,
\end{align}
where $S$ is a generic system coupling operator, and $t_{kq} = t_{qk}^*$ describes the reservoir-internal scattering~\footnote{
A toy model for such a reservoir could be a chain with $N$ modes (where the actual reservoir limit is obtained for $N\to\infty$) $H_B = \epsilon \sum_{i=1}^N d_i^\dagger d_i + \tau \sum_{i=1}^{N-1} (d_i^\dagger d_{i+1} + d_{i+1}^\dagger d_i)$ with homogeneous on-site energies $\epsilon$ and tunnelling amplitudes $\tau$. 
For this, one can use the unitary transformation $d_i = \sum_k \sqrt{2/(N+1)} \sin[\pi i k/(N+1)] c_k$ to diagonalize it exactly into
$H_B = \sum_{k=1}^N  \epsilon_k c_k^\dagger c_k$ with $\epsilon_k=\epsilon+2\tau\cos[\pi k/(N+1)]$.
When we couple this reservoir at its first site to a system via $H_I = S \otimes [d_1^\dagger d_1 - \langle d_1^\dagger d_1\rangle]$, we would obtain $t_{kq} = 2/(N+1) \cdot \sin[\pi k/(N+1)]\sin[\pi q/(N+1)]$.}
and the term with the Fermi function $f(\epsilon_k) = [e^{\beta(\epsilon_k-\mu)}+1]^{-1}$ has been introduced to cancel the expectation value of the coupling operator at equilibrium~\footnote{Without renormalization, such an interaction would perturb the equilibrium state of the reservoir.}.
The correlation function $C(\tau) = \traceB{e^{+\ii H_B\tau} B e^{-\ii H_B \tau} B \bar\rho_B}$ can then be written 
as a two-fold integral
$C(\tau) = \frac{1}{2\pi} \int d\omega \int d\omega' T(\omega,\omega') f(\omega) [1-f(\omega')] e^{\ii(\omega-\omega')\tau}$,
where we have introduced the function $T(\omega,\omega')=2\pi \sum_{kq} \abs{t_{kq}}^2 \delta(\omega-\epsilon_k)\delta(\omega'-\epsilon_q)=T(\omega',\omega)$ that describes the scattering processes inside the reservoir~\footnote{
For the chain model reservoir it would for $N\to\infty$ assume the form $T(\omega,\omega') = 2/(\pi \tau^2) \cdot \sqrt{1-(\omega-\epsilon)^2/(4\tau^2)}\sqrt{1-(\omega'-\epsilon)^2/(4\tau^2)} \Theta(4\tau^2-(\omega-\epsilon)^2)\Theta(4\tau^2-(\omega'-\epsilon)^2)$, which shows that by tuning $\epsilon$ and $\tau$ one may modify the spectral function of the reservoir.}.
Independent of the chemical potential $\mu$ (which may be used to tune the overall magnitude), the Fourier transform
$\gamma(\Omega) \equiv \int C(\tau) e^{+\ii\Omega\tau} d\tau = \int d\omega T(\omega,\omega+\Omega) f(\omega) [1-f(\omega+\Omega)]$
obeys the KMS relation $\gamma(-\Omega) = \gamma(+\Omega) e^{-\beta\Omega}$, which implies that the associated rates $R_{ab} = \gamma(E_b-E_a) \abs{\bra{a} S \ket{b}}^2$
obey the detailed balance condition~\eqref{EQ:local_detailed_balance} with $\mu_\nu\to 0$ and $\beta_\nu\to\beta$.
In order to have not too small rates, the actual chemical potential should be tuned near the desired transition energy.
Furthermore, energetic filtering for the chain toy model is possible by choosing $\epsilon$ also resonant with the desired transition energy and small $\tau$.
Further simplifications are possible when the scattering function is approximately constant~\footnote{When $T(\omega,\omega+\Omega)\approx T(\mu,\mu)\equiv T_\mu$, we obtain $\gamma(\Omega) \to T_\mu \Omega [1+n_B(\Omega)]$ with Bose distribution $n_B(\Omega) = [e^{\beta\Omega}-1]^{-1}$, leading to a transition rate 
$R_{ab} = T_\mu \abs{\bra{a} S \ket{b}}^2 (E_b-E_a)\left[1+n_B(E_b-E_a)\right]$.
}.

Unfortunately, particularly in the interesting Maxwell demon limit where $\delta\to \Delta$ one has two near-degenerate excitation energies in the system, and energetic filtering alone is insufficient.  
In this case, one can additionally tune the system and reservoir couplings to maintain the selectivity of the reservoirs~\footnote{
For a homogeneous triple quantum dot $H_S = \epsilon_S [d_1^\dagger d_1+d_2^\dagger d_2+d_3^\dagger d_3] - \tau_S [d_1^\dagger d_2 + d_2^\dagger d_3 + d_3^\dagger d_1 + {\rm h.c.}]$
one finds with $d_1 = (c_1-\sqrt{2} c_3)/\sqrt{3}$, $d_{2,3} = (\sqrt{2} c_1\mp \sqrt{3} c_2+c_3)/\sqrt{6}$ a representation in terms of two degenerate eigenmodes $H_S = [\epsilon_S-2\tau_S] c_1^\dagger c_1 + [\epsilon_S + \tau_S] [c_2^\dagger c_2 + c_3^\dagger c_3]$.
Then, (up to dephasing terms $\propto c_i^\dagger c_i$) the capacitive coupling $S_c \propto d_1^\dagger d_1$ drives transitions between states $1\leftrightarrow 3$, 
$S_h \propto d_1^\dagger d_1 + 2 d_3^\dagger d_3$ drives transitions between $1\leftrightarrow 2$ (desired) and $2\leftrightarrow 3$ (can be filtered), 
and finally $S_d \propto d_2^\dagger d_2$ drives all possible transitions (but $1\leftrightarrow 2$ and $1 \leftrightarrow 3$ can be filtered).
This can be further improved when non-homogeneous tripledot systems are considered.
}.

\section{Conclusions}

To answer the main question: Maxwell's demon, incarnated by autonomous demonic refrigerators, can be be pretty small (our considerations suggested three levels).
In addition, it does not require a bipartite system-controller setup and can also be very efficient in the sense that the entropy produced in the feedback loop is equivalent to the information current entering the effective controlled system.

The present contribution has focused on the mathematical framework of rate equations with the associated limitations.
Generalizations to equations describing situations of near-degenerate energies~\cite{trushechkin2021a}, with system-reservoir correlations~\cite{beyer2019a}, and the strong-coupling regime~\cite{strasberg2018a} in autonomous systems or not perfectly-energetic measurements~\cite{schaller2018a,seah2020a} and delay effects~\cite{debiossac2020a} in external loops are interesting pathways of further research.

\begin{acknowledgement}
The author thanks C. Kohlf\"urst and P. Strasberg for helpful comments on the manuscript and for generous conference support by the CRC 1242 (Project ID No. 278162697).
\end{acknowledgement}

\bibliographystyle{abbrv}
\bibliography{/home/schall96/literatur/postdoc/postdoc}

\end{document}